\documentclass{svjour2}                    
\smartqed  
\usepackage{graphicx}
\begin{document}

\title{Renormalization of competing interactions and superconductivity on small scales
}


\author{A.  Aharony, O. Entin-Wohlman, Y. Imry}


\institute{A. Aharony\at
Department of Physics, Ben Gurion University, \\
          Beer Sheva 84105, Israel \\
Raymond and Beverly Sackler School of Physics and Astronomy, \\
Tel Aviv University, Tel Aviv 69978, Israel\\
              Tel.: +972-8-6477558\\
              Fax: +972-8-6477008\\
              \email{aaharonyaa@gmail.com}      \\    
           \and
           O. Entin-Wohlman\at
          Department of Physics, Ben Gurion University, \\
          Beer Sheva 84105, Israel \\
Raymond and Beverly Sackler School of Physics and Astronomy, \\
Tel Aviv University, Tel Aviv 69978, Israel\\
\and
           Y. Imry \at
          Department of Condensed Matter Physics, Weizmann Institute of Science, \\
          Rehovot 76100, Israel}

\date{Received: date / Accepted: date}

\maketitle

\begin{abstract}
The interaction-induced orbital magnetic response of a nanoscale ring is evaluated for a diffusive system which is a superconductor in the bulk.
The interplay of the renormalized Coulomb and Fr\"{o}hlich interactions is crucial. The magnetic susceptibility which results from the fluctuations of the uniform superconducting order parameter is diamagnetic (paramagnetic) when the renormalized combined interaction is attractive (repulsive). Above the transition temperature of the bulk the total magnetic susceptibility has contributions from many wave-vector- and (Matsubara) frequency-dependent order parameter fluctuations. Each of these contributions results from a different renormalization of the relevant coupling energy, when one integrates out the fermionic degrees of freedom. The total diamagnetic response of the large superconductor may become paramagnetic when the system's size decreases.

\end{abstract}

\section{Introduction}
\label{intro}

 This paper is devoted to the memory of Ken Wilson. AA was a post-doc with Michael Fisher at Cornell for two years, beginning in July 1972. In the fall of 1972, Ken gave a course on the renormalization group, covering the material which later appeared in Ref. \cite{WK}. Every lecture gave the participants more tools to work with this new technique, and AA's publications during that period were deeply influenced by this course. On several occasions, AA presented his work to Ken, and he still appreciates his comments and encouragement. Already at that time, Ken talked about using the renormalization group for a wide variety of other problems which involve many (length or energy) scales. One example concerned the Kondo problem, which Wilson solved using his numerical renormalization group \cite{wilson}, following Anderson's ``poor man's" renormalization \cite{pwa}. Those ideas partly motivated the research reported below.

Renormalization, often accomplished using the renormalization-group method,
is one of the basic concepts
in physics. It deals with the way various coupling constants (e.g.,  the electron charge or the Coulomb interaction) change as a function of the relevant  scale for  the given problem. 
The scale may be
the resolution at which the system is examined,
determined by its size and by the relevant energy for the process under consideration. Often, one knows the coupling constant's  ``bare value"
at a less accessible (e.g., a very small or a very large) scale, and what is relevant for experiments is the value  at a different, ``physical" scale; the coupling  on the latter scale  is then used for the relevant physics \cite{fro}. 
In this paper we consider mesoscopic systems, of linear size ${\cal L}$, and calculate the size-dependence of both the relevant coupling constant and a  physical quantity. The paper aims to check whether measurements of the magntic suceptibility can yield information on the net interaction.

The example which we consider concerns the orbital magnetic susceptibility
  of a mesoscopic diffusive normal-metal ring, notably the persistent current \cite{BIL}, which flows in such a ring in response to an Aharonov-Bohm flux $\Phi$  which penetrates it. (Note: this is not the susceptibility associated with the superconducting order parameter!). In a recent paper \cite{HBS} we found that  this susceptibility, which is dominated by  the superconducting fluctuations (above the bulk superconducting transition), can change from being diamagnetic to paramagnetic as ${\cal L}$ decreases, typically reaching a few nm scale (see below).
Earlier work on this response, by Ambegaokar and Eckern \cite{AE}(a), found, to first order in the screened Coulomb interaction, that the persistent current in such a system is proportional to the averaged interaction, and therefore it is paramagnetic for a repulsive interaction. In a later paper \cite{AE}(b), they considered  a simple attractive interaction at the Debye energy, and found a diamagnetic response.  Hence, one would expect the magnetic response to change sign when the interaction changes its sign, and this would open the possibility to learn about the interaction by measuring the persistent current. However, as discussed below, the magnetic susceptibility contains many contributions, and each has a different magnetic response and a different effective interaction, so that  the situation is more complicated.

In the theory of superconductivity there exist two competing interactions: the repulsive Coulomb interaction,   starting on the large, microscopic energy scale--typically the Fermi energy or bandwidth $E^{}_{\rm F}$, and the attractive phonon-induced interaction,  operative only below the much smaller Debye energy $\omega^{}_{\rm D}$.
By integrating over thin shells in momentum (or energy) space \cite{REN}, one obtains  
the well-known variation of the electron-electron interaction coupling $g$, being repulsive or attractive, from 
a high-energy scale $\widetilde\omega_{>}$ 
to a low one $\widetilde\omega_{<}$,
\begin{equation}
\frac{1}{g(\widetilde\omega^{}_{<})}=\frac{1}{g(\widetilde\omega^{}_{>})}+\log\Bigl(\frac{\widetilde\omega^{}_{>}}{\widetilde\omega^{}_{<}}\Bigr)\ .\label{RG}
\end{equation}
(We use
$\hbar=c=k_{\rm B}=1$.)
Notice that a repulsive/attractive interaction is ``renormalized downwards/upwards"  with decreasing energy scale $\widetilde\omega^{}_<$.
What makes superconductivity possible is that at $\omega_{\rm D}$ the renormalized repulsion is much  smaller than its value on the microscopic scale. 
At $\omega_{\rm D}$ the attraction may  win and then at lower energies the total interaction increases in absolute value, until it diverges at some small scale, 
the conventional  $T^{}_c$ of the given material.

Equation (\ref{RG}) has been mainly used to derive the transition temperature for ${\bf q}=\nu=0$, i.e. for the space- and time-independent mean-field superconducting order parameter (or gap), which ignores both thermal and quantum fluctuations \cite{LV}.
In this paper we present and employ quantitative results for a diffusive normal metallic ring,  whose size ${\cal L}$ exceeds the elastic mean-free path.
For a diffusive system, the  typical  energy scale 
for many physical mesoscopic observables
should be  
the Thouless energy, namely the inverse of the time it takes an electron to traverse the finite system,
  $E^{}_{\rm T}=D/{\cal L}^{2}$, where
$D$ is the diffusion coefficient \cite{book}. Below, we use this energy scale to represent the size-dependence of the magnetic response of the mesoscopic rings. Specifically, we derive
the
superconductive fluctuation-induced partition function, and  obtain the effect of the two competing interactions in the Cooper channel. We use this result to
study the  average scale-dependent persistent current  of a large ensemble of metallic  rings in the presence of these  interactions, in the diffusive regime.

Section \ref{H} reviews the derivation of the free energy \cite{we,HBS}. After the renormalization due to the integration over the fermionic degrees of freedom, and after truncating the action at the quadratic order in the order parameters, this free energy ends up being a sum over Gaussian models, associated with ``order parameters" which depend on a wavevector ${\bf q}$ and on a bosonic Matsubara frequency $\nu$. Each of these models involves a different interaction strength, which depends on  ${\bf q}$ and on $\nu$. Section \ref{rg} expresses these coupling coefficients in terms of the renormalization group. Each of these energies is renormalized by an equation similar to Eq. (\ref{RG}). Above the bulk transition temperature $T^{}_c$, where only the uniform order parameter (with ${\bf q=0}$ and $\nu=0$) orders, all of these ``order parameters" fluctuate and affect the persistent current.
Section \ref{sus} presents the explicit expressions for the various contributions to the magnetic susceptibility coming from the various order parameters. Indeed, the results confirm that the magnetic susceptibility may turn from diamagnetic to paramagnetic as the ring becomes smaller (i.e. when the Thouless energy increases). The exact place when this switch happens depends on several parameters.
Section \ref{sum} contains our conclusions.

\section{The Partition Function}
\label{H}

Our initial Hamiltonian consists of a single-particle part, $H_{0}({\bf r})=(-i\nabla-e{\bf A})^{2}/(2m)-\mu+u({\bf r})$, and a term describing {\em local} (because of screening) repulsion and attraction, of coupling constants $\widetilde{g}_{r}\equiv\lambda_{r} {\cal N}/V$ and $g_{a}\equiv\lambda_{a}{\cal N}/V$, respectively,
\begin{equation}
{\cal H}=\int d{\bf r} \Bigl (\sum_{\sigma}\psi^{\dagger}_{\sigma}({\bf r})H^{}_{0}({\bf r})\psi^{}_{\sigma}({\bf r})\nonumber\\
+(\lambda^{}_{r}-|\lambda^{}_{a}|)
\psi^{\dagger}_{\uparrow}({\bf r})\psi^{\dagger}_{\downarrow}({\bf r})\psi^{}_{\downarrow}({\bf r})\psi^{}_{\uparrow}({\bf r})\Bigr )
\ ,\label{HAM}
\end{equation}
where
${\cal N}$ is the density of states,   $V$  the system volume,
and
$\psi^{}_{\sigma}$ ($\psi^{\dagger}_{\sigma}$) destroys (creates) an electron of spin  component  $\sigma$ at ${\bf r}$.
Here, $\lambda^{}_{r}$ is the bare repulsion (before any renormalization), $\lambda^{}_a$ is the attractive interaction (active only at energies below the Debye energy), $\mu$ is the chemical potential,
$u({\bf r})$ is the disorder potential due to nonmagnetic impurities, and ${\bf A}$ is the  vector potential. Below we assume that the magnetic field is perpendicular to the thin planar ring, of circumference ${\cal L}$, and therefore
 $A=\Phi /{\cal L}$, where $\Phi$ is the flux penetrating the ring, which gives  rise to a persistent current.
As the magnetic fields needed to produce  such currents are rather small, the Zeeman interaction will be ignored.

 As explained in Ref. \cite{HBS}, the partition function, ${\cal Z}$, corresponding to the Hamiltonian (\ref{HAM}) is calculated by the method of Feynman path integrals, combined with the Grassmann algebra of many-body fermionic coherent states \cite{AS}. Introducing the bosonic fields $\phi({\bf r},\tau)$ and $\Delta ({\bf r},\tau)$
via
{\em two} Hubbard-Stratonovich transformations
one is able to integrate over the fermionic variables, and express the partition function in terms of the bosonic fields. In the present paper we discuss only temperatures which are high relative to $T^{}_c$, at which it is sufficient to expand the results to second order in the bosonic fields \cite{COM3}.  This yields the Gaussian model,
\begin{equation}
\label{Zm}
 {\cal Z}/{\cal Z}_{0}= \int D(\Delta,\Delta^{\ast}_{}) \int D(\phi,\phi^{\ast}_{})
\exp[-\widetilde {\cal S}]\ .
\end{equation}
The action $\widetilde{\cal S}$ is a sum over the wavevectors ${\bf q}$ and the bosonic Matsubara frequencies $\nu=2\pi \ell T$, where $\ell$ is an integer,
\begin{eqnarray}
 \widetilde {\cal S}&=\sum_{{\bf q},\nu} \Bigl (\beta {\cal N} |\Delta^{}_{{\bf q},\nu}|^{2}/|g^{}_{a}|
+\beta {\cal N}|\phi_{{\bf q},\nu}|^{2}/\widetilde{g}^{}_{r} \nonumber\\
\nonumber\\
&-\left(
\Delta_{{\bf q},\nu}+i\phi^{}_{{\bf q},\nu}\right)\left(
\Delta_{{\bf q},\nu}^{\ast}+i\phi_{{\bf q},\nu}^{\ast}\right)
 \widetilde\Pi_{{\bf q},\nu} \Bigr )\ .\label{Am}
\end{eqnarray}
Here,  ${\cal Z}_{0}$ denotes the partition function of  noninteracting electrons,
and $\beta =1/T$.
Equation (\ref{Zm}) represents the partition function due to superconducting fluctuations.

The last term in Eq. (\ref{Am}) renormalizes the ``bare" interactions which appear in the first two terms there. This renormlization is written in terms of the polarization $\widetilde{\Pi}$ 
averaged over the impurity disorder  \cite{AGD}. In a diffusive system, this polarization
is given as a sum over fermionic Matsubara frequencies, $\omega =\pi T(2m+1)$, where $m$ is an integer,
\begin{equation}
\widetilde{\Pi}^{}_{{\bf q},\nu}
 = {\cal N}\sum_{\omega}
\frac{2\pi\Theta[\omega (\omega +\nu)]}{|2\omega+\nu|+Dq^{2}}=\frac{ {\cal N}}{T}\sum_{n}^{} \frac{1}{ n+ F^{}_{{\bf q},\nu} }  \ ,
\label{POL1}
\end{equation}
with
\begin{equation}
\label{F}
 \hspace{1cm}F^{}_{{\bf q},\nu}=0.5+[D({\bf q}-e{\bf A})^2+|\nu|]/(4\pi T)\  ,
\end{equation}
where $\Theta $ is the Heaviside function.
As discussed in the literature, Eq. (\ref{POL1}) represents a divergent sum
and therefore a cutoff is needed.
The cutoff energy is determined by the relevant interaction.
The sums in the terms of Eq. (\ref{Am})  involving $\Delta$ are bounded by the Debye energy $\omega_{\rm D}$ and therefore are cut off at
$m_{\max }=\omega^{}_{\rm D}/(2\pi T)-1$; these sums are denoted
$\Pi^{\omega_{\rm D}}$.   The sum multiplying $|\phi|^2$ is cut off by the bandwidth, which is of the order of the Fermi energy $E^{}_{{\rm F}}$, i.e., $m_{\max }=E^{}_{\rm F}/(2\pi T)-1$; this sum is denoted by $\Pi^{E_{\rm F}}$. Therefore, there appear two polarizations,
the one cut off  by $\omega=\omega_{\rm D}$ and the other by $\omega=E_{\rm F}$:
\begin{equation}
\frac{T}{{\cal N}}\Pi^{\omega}_{{\bf q},\nu}=
\Psi\Bigl (\frac{\omega}{2\pi T}+F^{}_{{\bf q},\nu}\Bigr )-
\Psi\Bigl (F^{}_{{\bf q},\nu}\Bigr )\equiv G^\omega(F^{}_{{\bf q},\nu})\ ,\label{PID}
\end{equation}
where $\Psi$ is the digamma function. 

Performing the integration over the fields $\phi$ yields
\begin{equation}
\label{z5a}
\frac{\cal Z}{{\cal Z}_0}=  \prod_{{\bf q},\nu}  \int D(\Delta^{}_{{\bf q},\nu},\Delta^{\ast}_{{\bf q},\nu})
\exp\Bigl [- (\beta {\cal N}/|g^{}_a|)a(F^{}_{{\bf q},\nu}) |\Delta^{}_{{\bf q},\nu}|^2 \Bigr ] \ ,
\end{equation}
where
\begin{equation}\label{a}
a(F^{}_{{\bf q},\nu})=g^{-1}_r({\bf q},\nu)\bigl(1+[g^{}_r({\bf q},\nu)-|g^{}_a|]G^{\omega^{}_{\rm D}}(F^{}_{{\bf q},\nu})\bigr)\ ,
 \end{equation}
while
$g_r({\bf q},\nu)$ is the renormalized wavevector- and frequency-dependent repulsion,
\begin{equation}
\label{gtil}
g^{-1}_{r}({\bf q},\nu)\
 =\widetilde{g}^{-1}_{r}+\Psi\Bigl (\frac{E^{}_{\rm F}}{2\pi T}+F^{}_{{\bf q},\nu}\Bigr )-\Psi\Bigl (\frac{\omega^{}_{\rm D}}{2\pi T}+F^{}_{{\bf q},\nu}\Bigr )\ .
\end{equation}
$g_r({\bf q},\nu)$ represents the effective repulsive interaction at energy $\omega=\omega^{}_{\rm D}$, after one integrates out the fermionic degrees of freedom going down in energy from $E^{}_{\rm F}$ to $\omega^{}_{\rm D}$. As can be seen from Eq. (\ref{gtil}), one has $0<g_r({\bf q},\nu)<\widetilde{g}^{}_r$; the renormalization weakens the repulsive interaction.
Since we are interested in the range $T^{}_c<<T<<\omega^{}_{\rm D}$, Eq. (\ref{gtil}) can be approximated by
\begin{equation}\label{gtil1}
 g^{-1}_{r}({\bf q},\nu)\
 =\widetilde{g}^{-1}_{r}+\log\Bigl (\frac{E^{}_{\rm F}+2\pi TF^{}_{{\bf q},\nu}}{\omega^{}_{\rm D}+2\pi TF^{}_{{\bf q},\nu}}\Bigr )\ .
\end{equation}
In the limit ${\bf q}=\nu=0$ this becomes $ g^{-1}_{r}({\bf q},\nu)\
 =\widetilde{g}^{-1}_{r}+\log\bigl (E^{}_{\rm F}/\omega^{}_{\rm D}\bigr)$, which is a special case of Eq. (\ref{RG}), with $\omega^{}_>=F^{}_{\rm F}$ and $\omega^{}_<=\omega^{}_{\rm D}$.

 Carrying out the integrations  over the fields $\Delta$ in Eq. (\ref{z5a})  yields (apart from trivial multiplicative constants)
\begin{equation}
{\cal Z}/{\cal Z}^{}_{0}=\prod_{{\bf q},\nu}a^{-1}_{{\bf q},\nu}\ ,\label{ZP}
\end{equation}
and therefore
the free energy  is a sum over Gaussian free energies of the ${\bf q}-$ and $\nu-$dependent ``order parameters" $\Delta^{}_{{\bf q},\nu}$,
 \begin{equation}
 {\cal F}={\cal F}^{}_0+T\sum_{{\bf q},\nu)} \log[a(F^{}_{{\bf q},\nu}]\ .
  \end{equation}
 This approximation remains valid as long as all $a$'s remain positive and relatively large \cite{COM3}. At very high temperatures, one has $a(F^{}_{{\bf q},\nu})\rightarrow \widetilde{g}^{-1}_r>0$ and all the ``order parameters" remain zero. As the temperature is lowered, $a(F^{}_{{\bf q},\nu})$ decreases. Upon cooling, if this coefficient crosses the value zero then one has
a phase transition of the corresponding ``order parameter" (and then one must include higher order terms in the action). The uniform order parameter $\Delta^{}_{{\bf 0},0}$ orders when  $a(F^{}_{{\bf 0},0})=0$, and all the other $a(F^{}_{{\bf q},\nu})$'s vanish at lower transition temperatures (if they vanish at all).
Close to the uniform transition, the physical properties are dominated by $a(F^{}_{{\bf 0},0})$. However, as we show below,  at temperatures which are significantly higher than all these critical temperatures the physical properties have significant contributions from the fluctuations of {\it all} the ``order parameters", and not only from those of the uniform one.

\section{The renormalization group}
\label{rg}

Equation (\ref{gtil1})
 is the same as Eq. (\ref{RG}), with $\widetilde\omega^{}_>=E^{}_{\rm F}+2\pi T F^{}_{{\bf q},\nu}$ and $\widetilde\omega^{}_<=\omega^{}_{\rm D}+2\pi T F^{}_{{\bf q},\nu}$.  In the spirit of Wilson's renormalization group, and of Anderson's ``poor man's scaling", one divides the fermionic energy axis into logarithmic segments, and then one sums over these segments \cite{wilson,pwa}. In our case, this logarithmic division is done by writing
 $\widetilde{\omega}={\omega+2\pi T F^{}_{{\bf q},\nu}}=(E^{}_{\rm F}+2\pi T F^{}_{{\bf q},\nu})e^{-l}$, where $l$ is the iteration index. It is then easy to see that Eq. (\ref{gtil1}) is the solution of the differential equation
 \begin{equation}\label{RGG}
 \frac{dg^{-1}}{dl}=1\ ,
 \end{equation}
 with $0<l<l^{}_1\equiv\log(\widetilde\omega^{}_>/\widetilde\omega^{}_<)=\log[(E^{}_{\rm F}+2\pi T F^{}_{{\bf q},\nu})/(\omega^{}_{\rm D}+2\pi T F^{}_{{\bf q},\nu})]$.

 At $\omega^{}_{\rm D}$, the attractive interaction starts to act, and the interaction $g_r({\bf q},\nu)$ is replaced by $g_r({\bf q},\nu)-|g^{}_a|$. This difference indeed appears in Eq. (\ref{a}).
 Since $G^\omega(F^{}_{{\bf q},\nu})>0$, $a$ can vanish only if one has $(g^{}_r({\bf q},\nu)-|g^{}_a|)<0$, i.e. when the effective interaction at $\omega=\omega^{}_{\rm D}$ is attractive. The renormalization of the repulsive interaction between $E^{}_{\rm F}$ and $\omega^{}_{\rm D}$ reduces the net coupling coefficient at $\omega^{}_{\rm D}$, and thus enhances the possibility of ordering. To formulate the results in the renormalization group language, we define an effective attractive interaction, $g^{}_{\rm eff}({\bf q},\nu)$, via
 \begin{equation}\label{geff}
\frac{1}{g^{}_{\rm eff}({\bf q},\nu)}\equiv\frac{1}{g^{}_r({\bf q},\nu)-|g^{}_a|}+\Psi\bigl(\frac{\omega^{}_{\rm D}}{2\pi T}+F^{}_{{\bf q},\nu}\bigr)-\Psi(F^{}_{{\bf q},\nu})\ .
 \end{equation}
 This expression looks exactly like Eq. (\ref{gtil}), and it represents the renormalization of the ``new' effective coupling coefficient, $g^{}_{\rm eff}$.  At large $F^{}_{{\bf q},\nu}$, one can again replace $\Psi(z)$ by $\log z$, and consider a gradual integration of the fermionic degrees of freedom down from $\omega^{}_{\rm D}$ to $\omega$,
\begin{equation}\label{gtil2}
 \frac{1}{g^{}_{r}({\bf q},\nu)}
 =\frac{1}{\widetilde{g}^{}_{r}}+\log\Bigl (\frac{\widetilde\omega^{}_{\rm D}}{\widetilde\omega}\Bigr )\ .
\end{equation}
 Using the same logarithmic division as above, this equation also looks like Eq. (\ref{RG}), and it can be interpreted as the solution of Eq. (\ref{RGG})
from its value at $l^{}_1$, $g^{}_{\rm eff}=g^{}_r({\bf q},\nu)-|g^{}_a|$, through intermediate values of $l$, to  $l^{}_2=\log[(E^{}_{\rm F}+2\pi T F^{}_{{\bf q},\nu})/(2\pi T F^{}_{{\bf q},\nu})]$. Interestingly, the upper limit of the iteration variable $l$ (and therefore also the lower limit of the renormalization in energy, $\omega^{}_<$), depends on the temperature $T$ and on the parameters ${\bf q}$ and $\nu$. In particular, for finite systems this upper bound depends on the system size via the discrete values of the momenta; for large $E^{}_{\rm T}$, one has $F^{}_{{\bf q},\nu}\sim Dq^2/(4\pi T) \sim E^{}_{\rm T}/(4\pi T)$, and the lower bound is of the order of the relevant energy scale $\omega^{}_< \sim E^{}_{\rm T}$.
Larger values of $|{\bf q}|$ and of $|\nu|$ imply larger values of $F^{}_{{\bf q},\nu}$, hence smaller values of $l^{}_2$. In particular, such larger values result is smaller effective interactions, and therefore in smaller transition temperatures (if at all).

 When $F^{}_{{\bf q},\nu}$ is not large (e.g. for ${\bf q}=\nu=0$), one cannot use the logarithmic approximation for $\Psi(z)$. In that case, the renormalization of $g^{-1}_{\rm eff}$ is given by
 \begin{equation}
\frac{1}{g^{}_{\rm eff}(\widetilde\omega^{}_<)}=\frac{1}{g^{}_{\rm eff}(\widetilde\omega^{}_>)}+\Psi\Bigl(\frac{\widetilde\omega^{}_>}{2\pi T}\Bigr)-\Psi\Bigl(\frac{\widetilde\omega^{}_<}{2\pi T}\Bigr)\ .
 \end{equation}
When $g^{}_{\rm eff}(\widetilde\omega^{}_>)<0$, it is easy to see that $g^{}_{\rm eff}$ becomes more negative as  $\widetilde\omega^{}_<$ decreases. From Eq. (\ref{geff}), one sees that the lower bound of the renormalization iterations (the last term there) corresponds to the physical value $\widetilde\omega^{}_<=2\pi T F^{}_{{\bf q},\nu}$.
 The vanishing of $a(F^{}_{{\bf q},\nu})$ is  equivalent to the divergence of $g^{}_{\rm eff}({\bf q},\nu)$, and the corresponding transition temperature is given by the solution of the equation $g^{-1}_{\rm eff}({\bf q},\nu)=0$. In particular, since $\omega^{}_{\rm D}>>T$, the ``uniform" transition temperature $T^{}_c$, for the ordering of $\Delta^{}_{{\bf 0},0}$, is the solution of the equation $1+[g^{}_r({\bf 0},0)-|g^{}_a|]\bigl(\bigl[\log[\omega^{}_{\rm D}/(2\pi T^{}_c)]-\Psi(1/2)\bigr]\bigr)=0$, reproducing the ``usual" BCS transition temperature for the bulk,  $T^{}_c=\omega^{}_{\rm D}\exp\{\Psi[1/2]-1/[|g^{}_a|-g^{}_r({\bf 0},0)]\}/(2\pi)$ \cite{LV}.

\section{The magnetic susceptibility}
\label{sus}

We now consider a thin ring, of circumference ${\cal L}$. The shifted longitudinal momentum component, tangential to the ring, is
given by $q^{}_{||}=2\pi(n+2\phi)/{\cal L}$, where $n$ is an integer and where $\phi=\Phi/\Phi^{}_0=HS/\Phi^{}_0$, $\Phi^{}_0=h/e$ being the flux quantum and $S={\cal L}^2/(4\pi)$ being the area of the ring \cite{we}. For a thin ring we ignore transverse momenta, and Eq. (\ref{F}) becomes
\begin{equation}
\label{F1}
 F^{}_{{\bf q},\nu}=\frac{1+\ell}{2}+\frac{\pi E^{}_{\rm T}}{T}(n+2\phi)^2\ ,
\end{equation}
with the Thouless energy  $E^{}_{\rm T}\equiv D/{\cal L}^2$.  The orbital magnetic susceptibility due to the superconduction fluctuations is given by
\begin{equation}\label{ch1}
\Delta\chi=-\frac{\partial^2{\cal F}}{\partial H^2}=-\overline{\chi}^{}_0\sum_{{\bf q},\nu}\frac{\partial^2}{\partial\phi^2}\log[a(F^{}_{{\bf q},\nu})]\ ,
\end{equation}
where $\overline{\chi}^{}_0=T[{\cal L}^2/(4\pi \Phi_0)]^2$. Using Eq. (\ref{F1}), the contribution of the fluctuations of $\Delta^{}_{{\bf q},\nu}$ is
\begin{equation}
\Delta\chi_{n,\ell}=-\overline{\chi}^{}_0\frac{\partial^2}{\partial\phi^2}\log[a(F^{}_{{\bf q},\nu})]=-4\overline{\chi}^{}_0\frac{\partial^2}{\partial n^2}\log[a(F^{}_{{\bf q},\nu})]
\ .
\end{equation}
Using the relation
\begin{equation}
\frac{\partial\log(a)}{\partial n}=\frac{\partial\log(a)}{\partial F}\frac{\partial F}{\partial n}=\frac{\partial\log(a)}{\partial F}\frac{2\pi E^{}_T(n+2\phi)}{T}\ ,
\end{equation}
 we also have
\begin{equation}\label{ch2}
\Delta\chi^{}_{n,\ell}=-\chi^{}_0\Bigl[\frac{\partial^2\log(a)}{\partial F^2}
\frac{2\pi E^{}_{\rm T}}{T}(n+2\phi)^2+\frac{\partial\log(a)}{\partial F}\Bigr]\ ,
\end{equation}
where $\chi^{}_0=4\overline{\chi}^{}_02\pi E^{}_{\rm T}/T=D{\cal L}^2/(2\pi \Phi^2_0)$.

\begin{figure*}
  \includegraphics[width=0.5\textwidth]{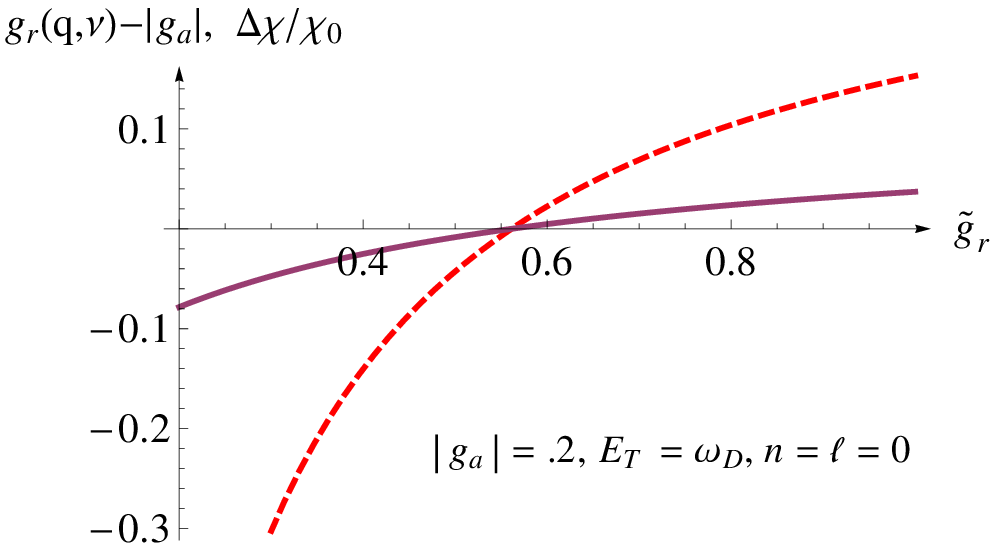}\ \ \ \includegraphics[width=0.5\textwidth]{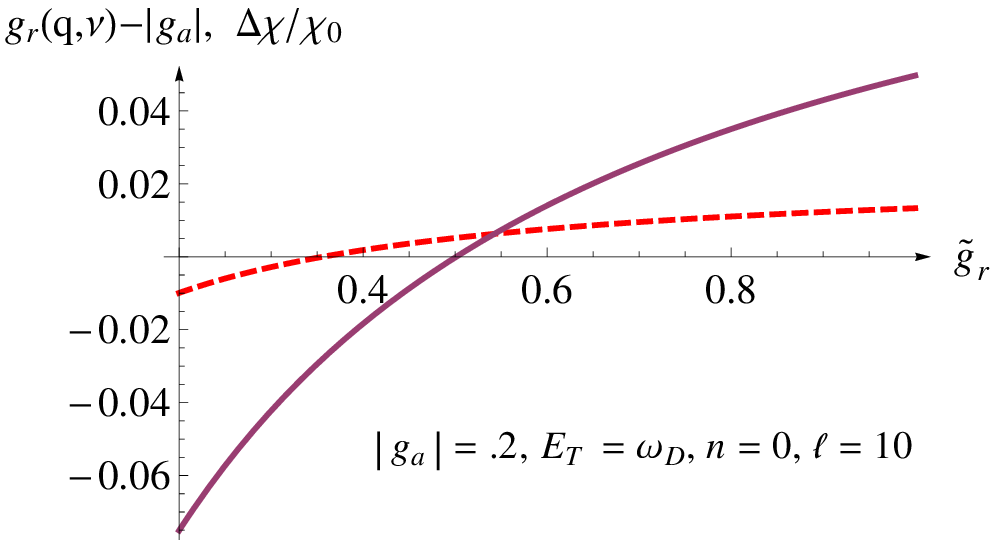}\\
  \includegraphics[width=0.5\textwidth]{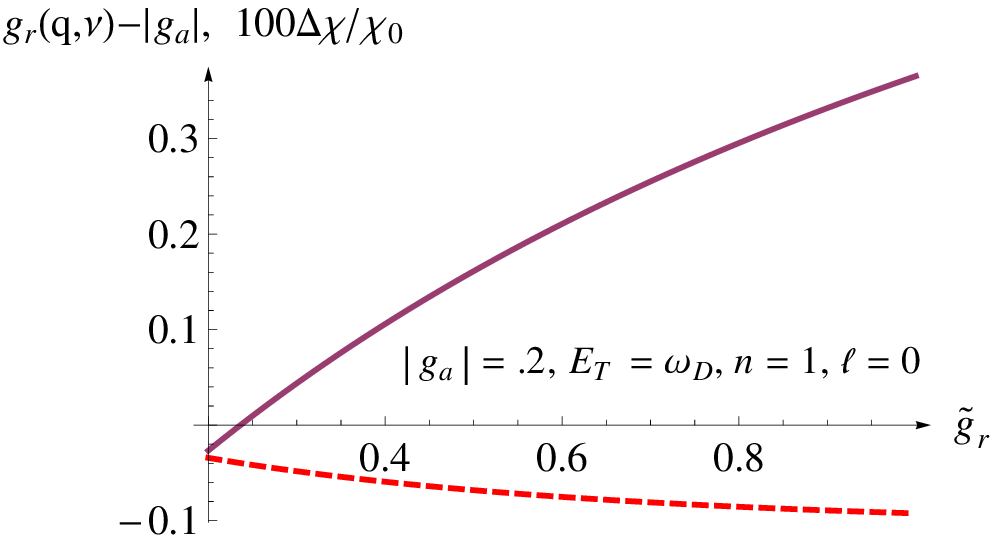}\ \ \ \includegraphics[width=0.5\textwidth]{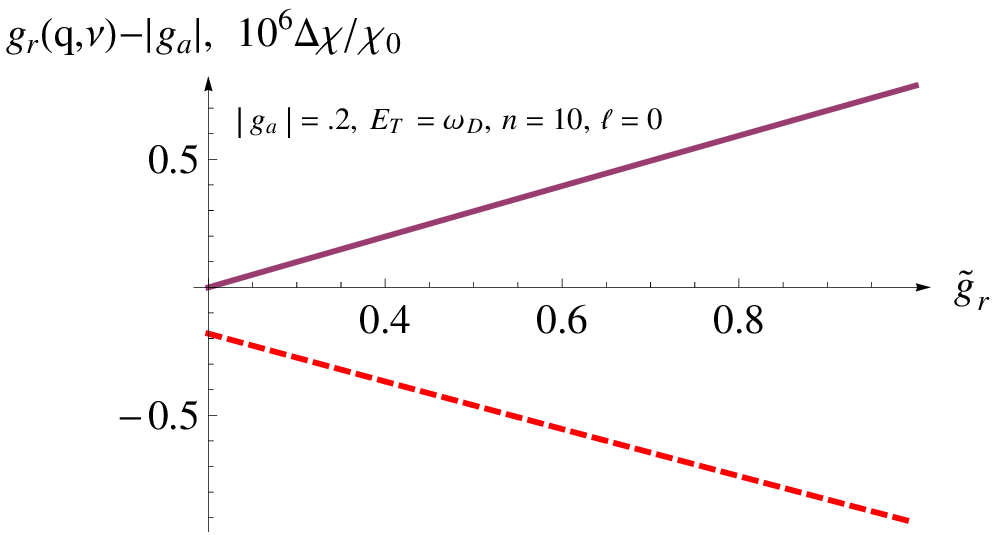}
\caption{The dependence of the renormalized effective interaction (full lines) and of  the corresponding contribution to the zero-field magnetic susceptibility (dashed lines) on the bare repulsive energy $\widetilde{g}^{}_r$, for several values of $n$ and $\ell$. The other parameters are  $\omega^{}_{\rm D}=125T/(2\pi)$, $E^{}_{\rm F}=25 \omega^{}_{\rm D}$, $E^{}_{\rm T}=\omega^{}_{\rm D}$.}
\label{fig1}
\end{figure*}

From now on we present results only for zero field, $\phi=0$. Figure \ref{fig1} shows the dependence of $\Delta\chi^{}_{n,\ell}$ on the bare interaction $\widetilde{g}^{}_r$. The figure also shows  the  the effective interaction at $\omega^{}_{\rm D}$,  $g_r({\bf q},\nu)-|g^{}_a|$, for the same parameters. As expected, the effective interaction becomes more attractive as $\widetilde{g}^{}_r$ decreases.
For $n=0$, $\Delta\chi^{}_{0,\ell}$ also decreases as the effective interaction decreases. For $n=\phi=0$, the magnetic susceptibility is given by the second term in Eq. (\ref{ch2}). In this case, both the effective interaction and the contribution to the magnetic susceptibility change sign, from being repulsive and paramagnetic at high $\widetilde{g}^{}_r$ to being attractive and diamagnetic at low
$\widetilde{g}^{}_r$. This might be expected intuitively: a strong repulsive (attractive) interaction is expected to yield paramagnetic (diamagnetic) behavior. Indeed, such a change of sign of the magnetic response might be expected based on the papers by Ambegaokar and Eckern \cite{AE}. However, we find that the two curves cross zero at the same point only for the ``classical" case, $n=\ell=0$, and not for the other $n$'s and $\ell$'s. Therefore, it is not quantitatively true that the sign of the interaction determines the sign of the total magnetic susceptibility.

For $n\ne 0$, the contributions to the magnetic susceptibility {\it always} remain  diamagnetic. In this case, $\Delta\chi^{}_{n,\ell}$ is dominated by the first term in Eq. (\ref{ch2}), which is always negative. Furthermore, in this case $\Delta\chi^{}_{n,\ell}$ becomes {\it more} diamagnetic as the effective interaction is more repulsive! However, as seen from the scales in Fig. \ref{fig1}, the contributions to the magnetic susceptibility from these terms are very small in magnitude.

\begin{figure*}
  \includegraphics[width=0.5\textwidth]{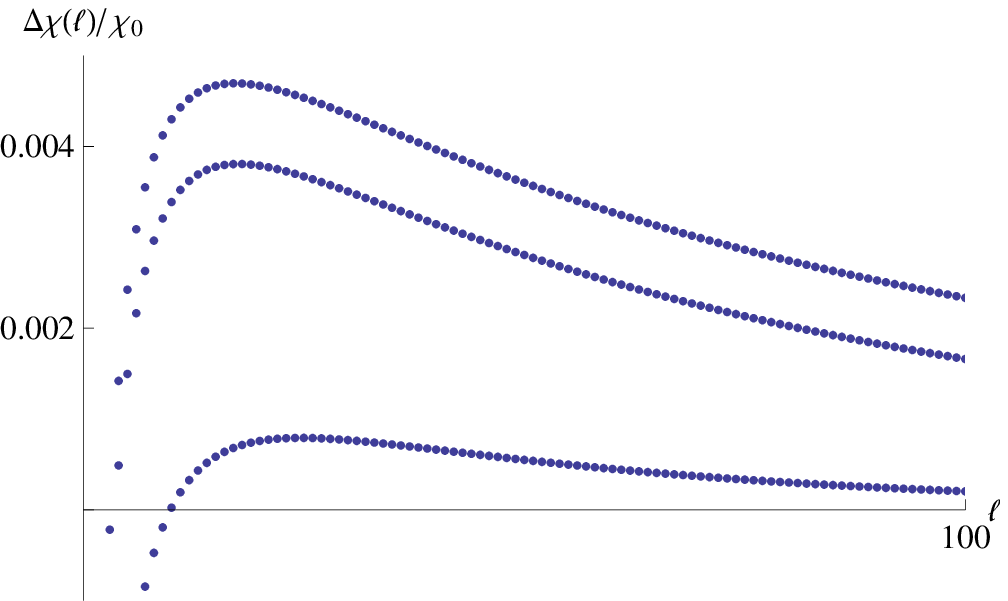}\ \ \ \includegraphics[width=0.5\textwidth]{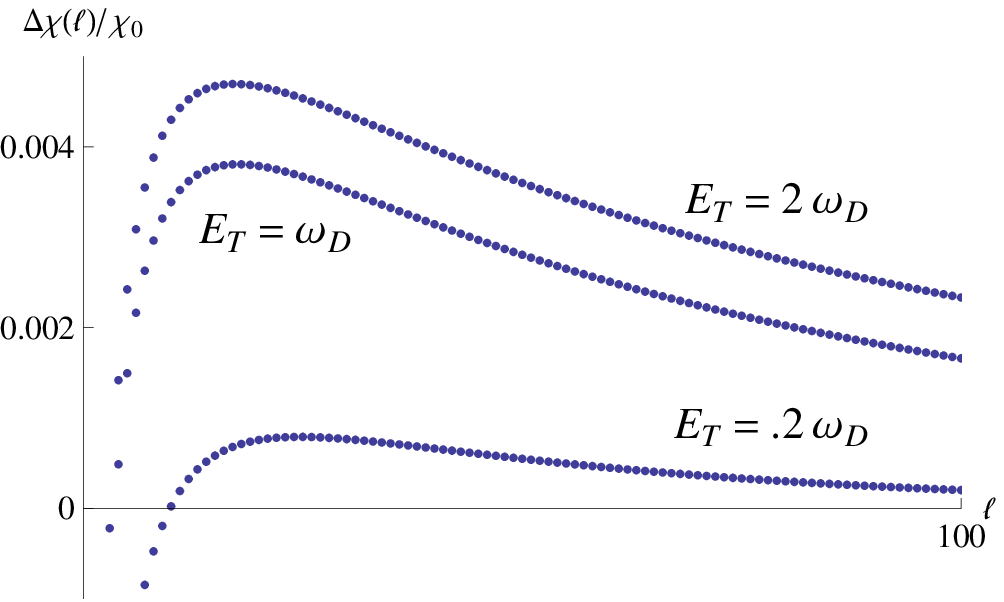}\\
  \includegraphics[width=0.5\textwidth]{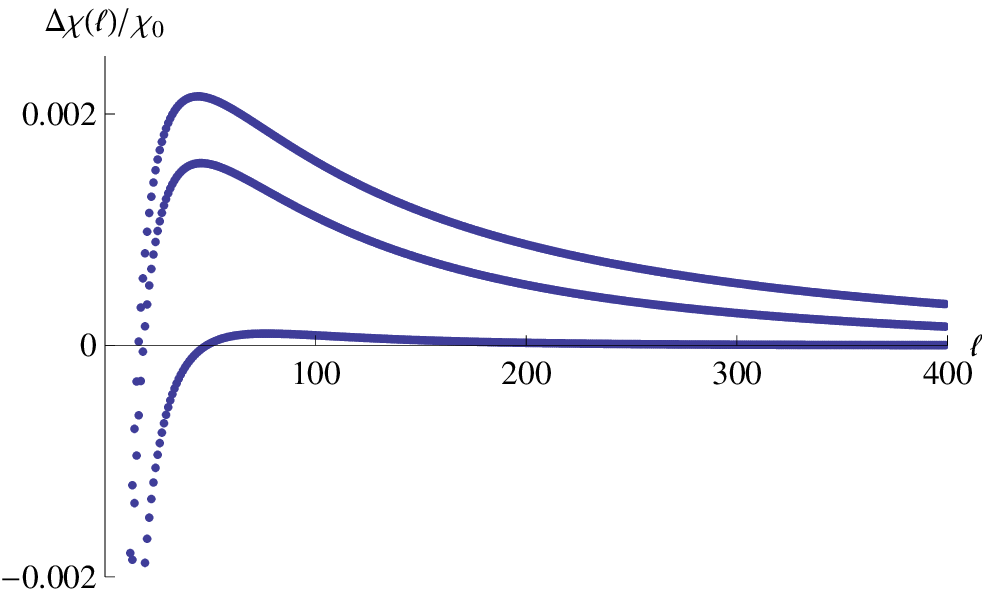}\ \ \ \includegraphics[width=0.5\textwidth]{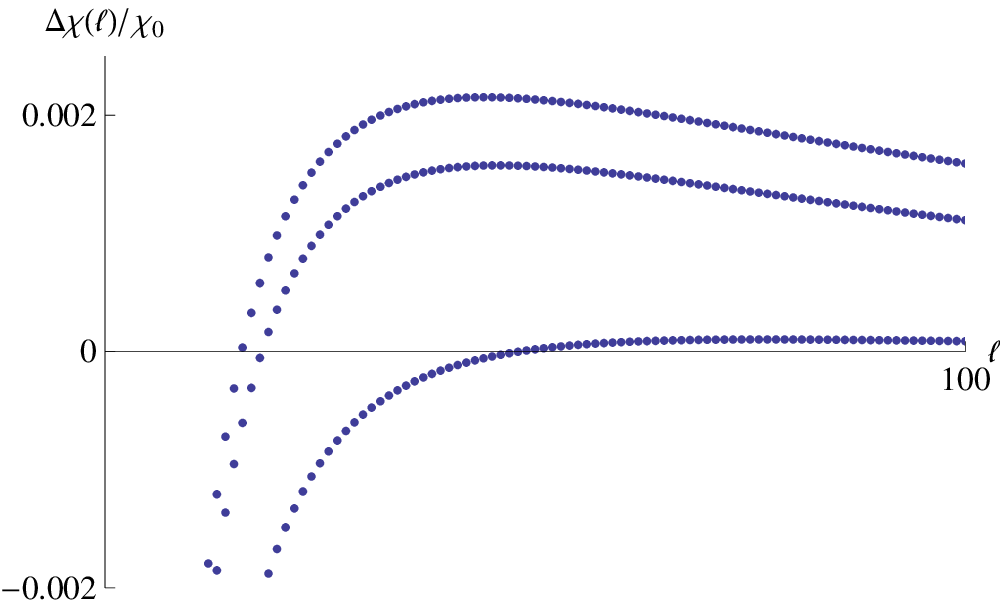}
\caption{The dependence of $\Delta\chi(\ell)$ on $\ell$, for several values of $E^{}_{\rm T}$. Top:  $\widetilde{g}_r=.5$. Bottom: $\widetilde{g}_r=.3$. The three lines on each plot correspond to the same three values of $E^{}_{\rm T}$, as indicated on the first plot. The other parameters are $|g^{}_a|=.2$, $\omega^{}_{\rm D}=125T$, $E^{}_{\rm F}=25 \omega^{}_{\rm D}$.}
\label{fig2}
\end{figure*}

To find to total magnetic susceptibility, one must sum over  the contributions from all the fluctuating order parameters, Eq. (\ref{ch1}). Since the contributions from $n\ne 0$ decay strongly with $n$, we first performed the summation over $n$. It turns out that after a few terms, one obtains excellent results by replacing the sum by an integral:
\begin{eqnarray}
&\Delta\chi(\ell)=\Delta\chi^{}_{0,\ell}+2\Bigl[\sum_{n=1}^{n_1-1}\frac{\partial^2\log(a)}
{\partial n^2}+I\Bigr]\ ,\nonumber\\
&I=\int_{n^{}_1}^\infty dn\frac{\partial^2\log(a)}{\partial n^2}=\frac{\log(a)}{\partial n}\Big|_{n=\infty}-\frac{\log(a)}{\partial n}\Big|_{n=n_1}\ .
\end{eqnarray}
In most of our calculations we used $n_1=20$. The sums usually already converged at this value, and the integral $I$ was negligible. The results do not change for larger values of $n_1$.
Figure \ref{fig2} shows results for $\Delta\chi(\ell)$. It is diamagnetic at small $\ell$, becomes paramagnetic at larger $\ell$ and decays to zero as $\ell\rightarrow\infty$. For a given value of $g^{}_a$, the crossing point from dia- to paramagnetic behavior depends on both $E^{}_{\rm T}$ and on $\widetilde{g}^{}_r$.

\begin{figure*}
  \includegraphics[width=0.5\textwidth]{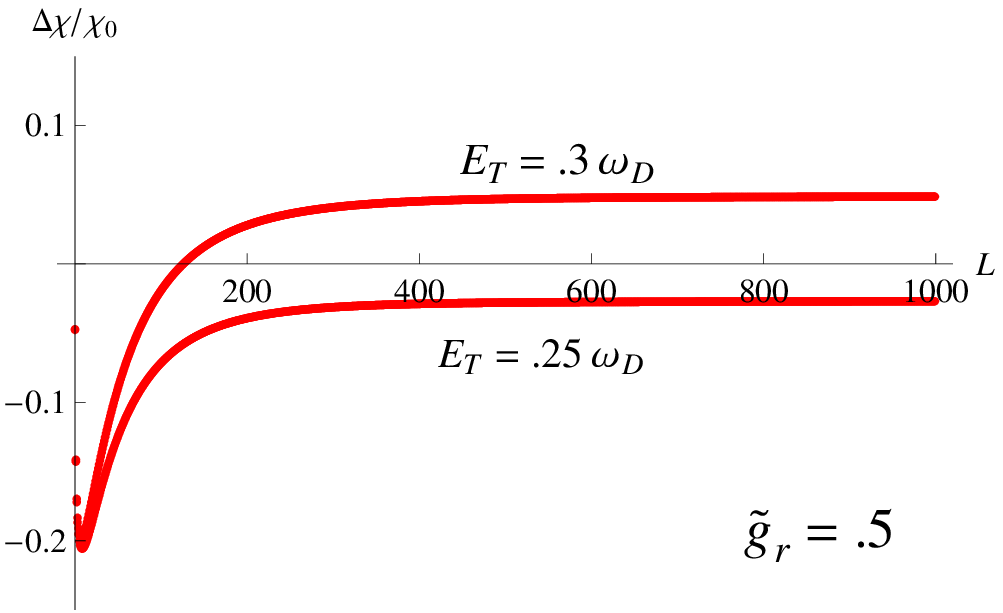}\ \ \ \includegraphics[width=0.5\textwidth]{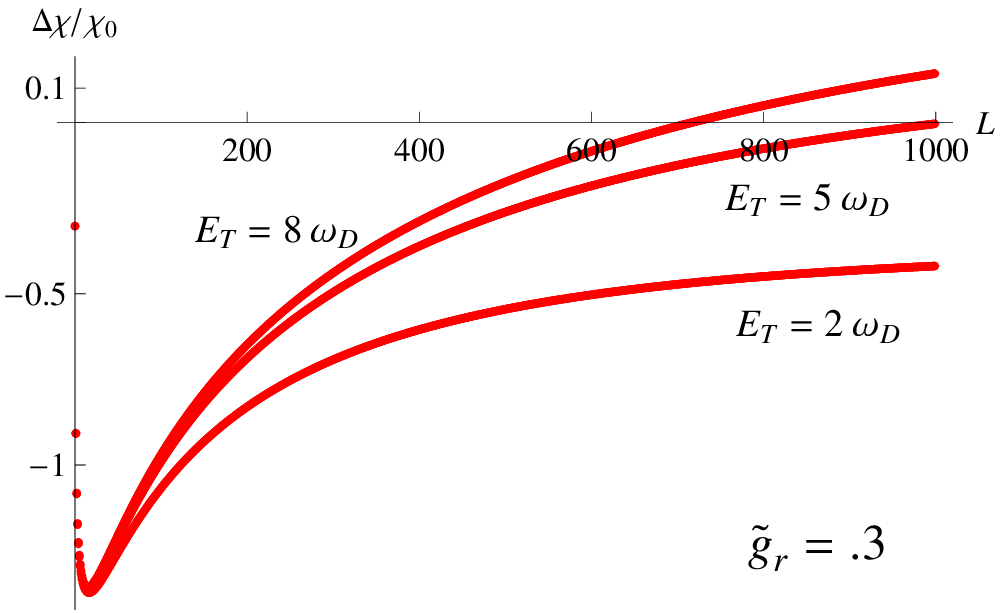}
\caption{The dependence of the total magnetic susceptibility $\Delta\chi$ on the cutoff $L$, for several values of $E^{}_{\rm T}$ and $\widetilde{g}_r$. The other parameters are $|g^{}_a|=.2$, $\omega^{}_{\rm D}=125T$, $E^{}_{\rm F}=25 \omega^{}_{\rm D}$.}
\label{fig3}
\end{figure*}

The decay of $\Delta\chi(\ell)$ to zero is slow, roughly as $1/\ell$, and therefore the total magnetic susceptibility,
\begin{equation}
\Delta\chi=\Delta\chi(0)+2\sum_{\ell=1}^L\Delta\chi(\ell)\ ,
\end{equation}
grows slowly (logarithmically) with the upper cutoff $L$. Figure \ref{fig3} shows several examples of the $L-$dependence of $\Delta\chi$. The qualitative results do not depend on $L$: for $|g^{}_a|=.2$ and $\widetilde{g}^{}_r=.5$, the total magnetic susceptibility changes sign at $E^{}_{\rm T}\approx .27\omega^{}_{\rm D}$. In contrast, when $\widetilde{g}^{}_r=.3$, the dependence on the cutoff is stronger, and this switch happens only at relatively large $E^{}_{\rm T}$, of order $20\omega^{}_{\rm D}$. Choosing an upper cutoff at $|\nu|<E^{}_{\rm F}$ implies $L=500$.

\section{Conclusions}\label{sum}

Our results confirm the expectation based on Ref. \cite{AE} for the zero field magnetic susceptibility which results from the fluctuations of the uniform order parameter, $\Delta\chi^{}_{0,0}$. This susceptibility indeed turns from diamagnetic to paramagnetic  exactly when the corresponding effective interaction,  $g^{}_{\rm eff}({\bf 0},0)$ changes from being attractive to being repulsive. However, within our Gaussian approximation this ``uniform" contribution to the magnetic susceptibility does not depend on the Thouless energy $E^{}_{\rm T}$, and therefore cannot be used to study the size-dependence of the effective interaction. Unlike possible intuitive expectations, the other individual contributions, $\Delta\chi^{}_{n,\ell}$, do not have any precise relation to the corresponding effective interactions. As a result, this is also true for the full magnetic susceptibility, $\Delta\chi$.

The total magnetic susceptibility, $\Delta\chi$, does depend on $E^{}_{\rm T}$, and therefore on the ring size ${\cal L}$. When the bare repulsive interaction $\widetilde{g}^{}_r$ is not too large, $\Delta\chi$ remains diamagnetic. However, as $\widetilde{g}^{}_r$ increases (for a given value of the attractive interaction $g^{}_a$), $\Delta\chi$ does become paramagnetic at large enough values of $E^{}_{\rm T}$, confirming the conclusions of Ref. \cite{HBS}. From our results, this happens when $E^{}_{\rm T}$ is of order $\omega^{}_{\rm D}$ (up to an order of magnitude one way or the other). In a dirty superconductor the transition temperature $T^{}_c$ is of order $D/\xi(0)^2$, where $\xi(0)$ is the Landau-Ginzburg coherence length. Also, typically the Debye energy $\omega^{}_{\rm D}$ is larger by two orders of magnitude than $T^{}_c$. Thus the relevant range of ${\cal L}$ is about an order of magnitude below $\xi(0)$, namely on the nm scale.

Although we find no direct relation between the signs of the effective interaction and of the magnetic susceptibility, such a relation still holds qualitatively \cite{HBS}. It would be interesting to see experimental confirmations of our main qualitative conclusion: the magnetic response may change sign as the ring becomes smaller.
 Whereas an overall total attraction suffices to lead to a superconducting phase in the bulk material, we find that it does  not ensure a diamagnetic response of the mesoscopic system. The reason being that in mesoscopic rings  the orbital  magnetic response owes its very existence to a finite Thouless energy. Therefore, when the latter is large enough it can cause the response to be paramagnetic, albeit the strength of the attractive interaction. As the Thouless energy can be controlled experimentally, one may hope that the prediction made in this paper will be put to an experimental test.  Relatively large values of persistent currents may be achieved in molecular systems, and small discs are expected to behave similarly at small fluxes.
Finally we remark that there may already be experimental indications
to the validity of our prediction. Reich {\it et al.} \cite{REICH} found that thin enough gold films are paramagnetic. This may well be due to a small grain structure. Similar results were reported for small metallic particles in Ref. \cite{para}. 
A systematic experimental study of the size-dependence of the magnetic susceptibility of metallic nanoparticles is thus called for.


\begin{acknowledgements}

We thank Yuval Oreg and Alexander Finkelstein for important discussions, and Hamutal Bary-Soroker for participation in Ref. \cite{HBS}, which led to the present work.
This work was supported by the Israeli Science Foundation (ISF) and the US-Israel Binational Science Foundation (BSF).
\end{acknowledgements}



\end{document}